\documentclass[twocolumn,showpacs,floatfix]{revtex4}
\usepackage{epsf,amssymb,mathptm}

\begin{document}
\draft
\title{Avalanche dynamics driven by adaptive rewirings in complex
networks}
\author{K. Rho, S. R. Hong, and B. Kahng}
%\affiliation{
\address{School of Physics and Center for Theoretical Physics,
Seoul National University, Seoul 151-747, Korea}
\date{\today}
\thispagestyle{empty}
\begin{abstract}
We introduce a toy model displaying the avalanche dynamics of
failure in scale-free networks. In the model, the network growth 
is based on the Barab\'asi and Albert model and each node is
assigned a capacity or tolerance, which is constant irrespective
of node index. The degree of each node increases over time.
When the degree of a node exceeds its capacity, it fails and
each link connected to it is is rewired to other unconnected 
nodes by following the preferential attachment rule. 
Such a rewiring edge may trigger another failure. This
dynamic process can occur successively, and it exhibits a
self-organized critical behavior in which the avalanche size
distribution follows a power law. The associated exponent is $\tau
\approx 2.6(1)$. The entire system breaks down when any rewired 
edges cannot locate target nodes: the time at which this occurs 
is referred to as the breaking time. We obtain the breaking time 
as a function of the capacity. Moreover, using extreme value 
statistics, we determine the distribution function of the breaking time. 
\end{abstract}
 \maketitle

Complex systems are composed of many constituents that interact with
each other or adapt to external perturbations~\cite{nature,science}. 
Recently, there have been increasing attempts to describe such 
systems in terms of networks~\cite{review1,review2,review3,review4}, 
where nodes and links represent constituents and their interactions, 
respectively.
Many complex networks in real systems follow a power-law
$P_d(k)\sim k^{-\gamma}$ in the degree distribution, where degree
is the number of links connected to a given node~\cite{ba}. Such a
network is called scale-free (SF) networks. SF networks are
ubiquitous in real world, whose examples include the Internet and
the world-wide web, the metabolic networks, the protein
interaction networks, the co-authorship networks, etc.

It was studied that complex networks are robust against the random
removal of nodes, however, they are vulnerable to the intentional
removal of nodes with high degree~\cite{attack}. More severe
damage can be caused by triggering a few nodes, but the
failure propagates to other nodes in a cascading manner. 
Avalanche dynamics occurs frequently in complex networks due to 
the small-world feature of the system. The blackout of power-supply 
network in the United States in 1996 and 2003 is a
typical example of such a cascading failure in complex
networks~\cite{blackout}. Internet traffic is another example. 
In October 1986, during the first documented
Internet congestion collapse, the speed of the connection between
the Lawrence Berkeley Laboratory and the University of California
at Berkeley, which are separated by only 200 meters, decreased by a
factor of 100~\cite{berkeley}. These are only a few
instances; many others can be found in various systems such 
as cultural fad, earthquake, etc. 
Since the subject of avalanche dynamics in
complex systems is interesting and intricate, it has been 
studied extensively~\cite{aval1,aval2,aval3,aval4,aval5,aval6}.

Here we introduce a toy model exhibiting the avalanche dynamics in
complex networks. In this model, when a node fails, the links 
connected to it are rewired to other nodes, thereby preserving 
the total number of links during the avalanche dynamics. 
Due to the dynamic rule,
the overload imposed on a specific node is shared among the other nodes
globally as in the case of global load sharing in a fiber bundle
model.

Let us begin with the introduction of the model we consider here.
The model is based on the Barab\'asi and Albert (BA)
model~\cite{ba}, in which at each time step, a new node is added
and its links are connected to $m$ distinct existing nodes by following the
so-called preferential attachment (PA) rule. The newly added node is
connected to node $i$ whose degree is $k_i$ with a probability $\Pi_i=m
k_i/\sum_j k_j$. In this case, the number of nodes at $t=0$ is taken as $m$
and they are fully connected to each other. Our model is modified from 
the BA model as follows: We assign a capacity
or tolerance denoted by $\sigma$ to each node $i$. It is constant
independent of the node index $i$. The capacity represents the 
maximum number of connections that can be sustained by a node. 
As time goes on, the degree of each node increases. 
When the degree of a node exceeds its capacity $\sigma$, it is 
considered to be overloaded and is deleted from the system. 
Then all the links connected to the failed node are rewired 
to the other remaining nodes by applying the PA rule. 
By other remaining nodes, we mean the nodes in the system except 
the overloaded nodes. Multiple connections are not allowed. 
The rewiring process, shown in Fig.~1, may trigger the avalanche 
dynamics: when the degree of a node that receives a rewired link 
of an overloaded node exceeds $\sigma$, i.e. it becomes $\sigma+1$, 
it fails and its links must be rewired again. 
This process repeats until all the overloaded nodes are eliminated.
Thus the degree of each node that remains after the completion 
of the avalanche process does not exceed the capacity $\sigma$.
In the absence of a target node to which a rewiring link should 
be connected, the dynamics process is terminated and the entire 
system is considered to be collapsed, and the time at which 
occurs is referred to as the breaking time denoted by $t_b$. 
It is noteworthy that the avalanche process does
not spread locally from the triggering node, but it occurs across 
the entire system. Moreover the number of links is preserved during 
this process since the load of the failed node is distributed 
to other nodes globally. Otherwise, the system breaks down.

\begin{figure}[t]
\centerline{\epsfxsize=8.5cm \epsfbox{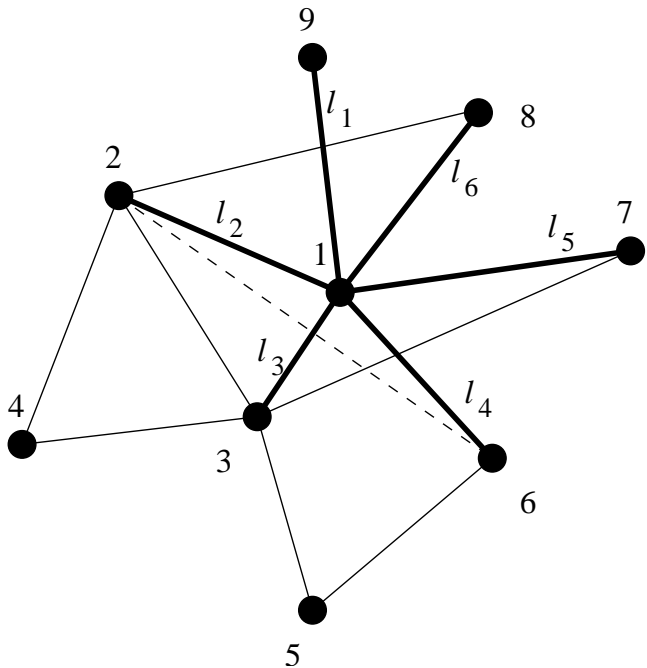}}
\caption{Illustration of the dynamic rule of the model for the
case of $\sigma=5$. When node 9 is newly added and connected to
node 1, the degree of node 1 exceeds its capacity $\sigma=5$. Then
the node 1 is overloaded and deleted from the system, and the
links $\ell_1 \sim \ell_6$ (bold lines) connected to it 
are rewired. For example, the link $\ell_2$ is rewired from 1 to
one of the nodes 5, 6, 7, or 9. For example, it is 
rewired to node 6, pivoted on node 2. The target
is selected according to the PA rule. } \label{critical_time}
\end{figure}

\begin{figure}[t]
\centerline{\epsfxsize=8.5cm \epsfbox{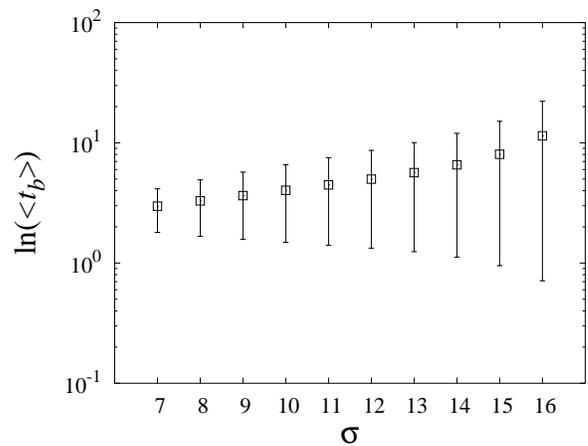}}
\caption{Plot
of the logarithm of the mean breaking time $\langle t_b \rangle$
as a function of capacity $\sigma$ on semi-logarithmic scale. The
linearly increasing behavior in this plot means 
$\langle t_b \rangle \sim \exp\exp
(\sigma)$. The data show that the mean breaking time seems 
to increase at a higher rate than the double exponential function.}
\label{critical_time}
\end{figure}

\begin{table}
\begin{center}
\begin{tabular}{r r r r r r}
\hline \hline $\sigma$ & $\langle t_b \rangle$ & $e$ &
$a$ & $b$ & $r$ \\
\hline
7  & 19 & 3 & 2.0& 34000 &3.2  \\
8 &  26 & 5& 2.8 &1900 &1.96\\
9 & 38 & 7 &2.68& 3074& 1.920\\
10 & 56 & 12& 2.68 &3470& 1.766\\
11 & 87 & 21 &2.57 & 4679 &1.668\\
12 & 147 & 38 &3.41 &1222 &1.165\\
13 &  283  &  81& 3.69 & 1106 &0.998\\
14 & 703 & 229 &4.05& 829& 0.803\\
15 & 3073 & 1188 &4.755 &512 &0.576\\
16 & 93075 & 45649& 6.412 &302 &0.334\\
\hline \hline
\end{tabular}
\end{center}
\caption{Summary table : capacity ($\sigma$), the mean
breaking time ($\langle t_b \rangle$), the root-mean-square of the
breaking time ($e$) and the estimated values of the constants $a$,
$b$, and $r$ in the probability distribution function
$F(t)=1-\exp[-\exp(a-b {t}^{-r})]$} \label{table:parameter}
\end{table}

\begin{figure}
\centerline{\epsfxsize=8.5cm \epsfbox{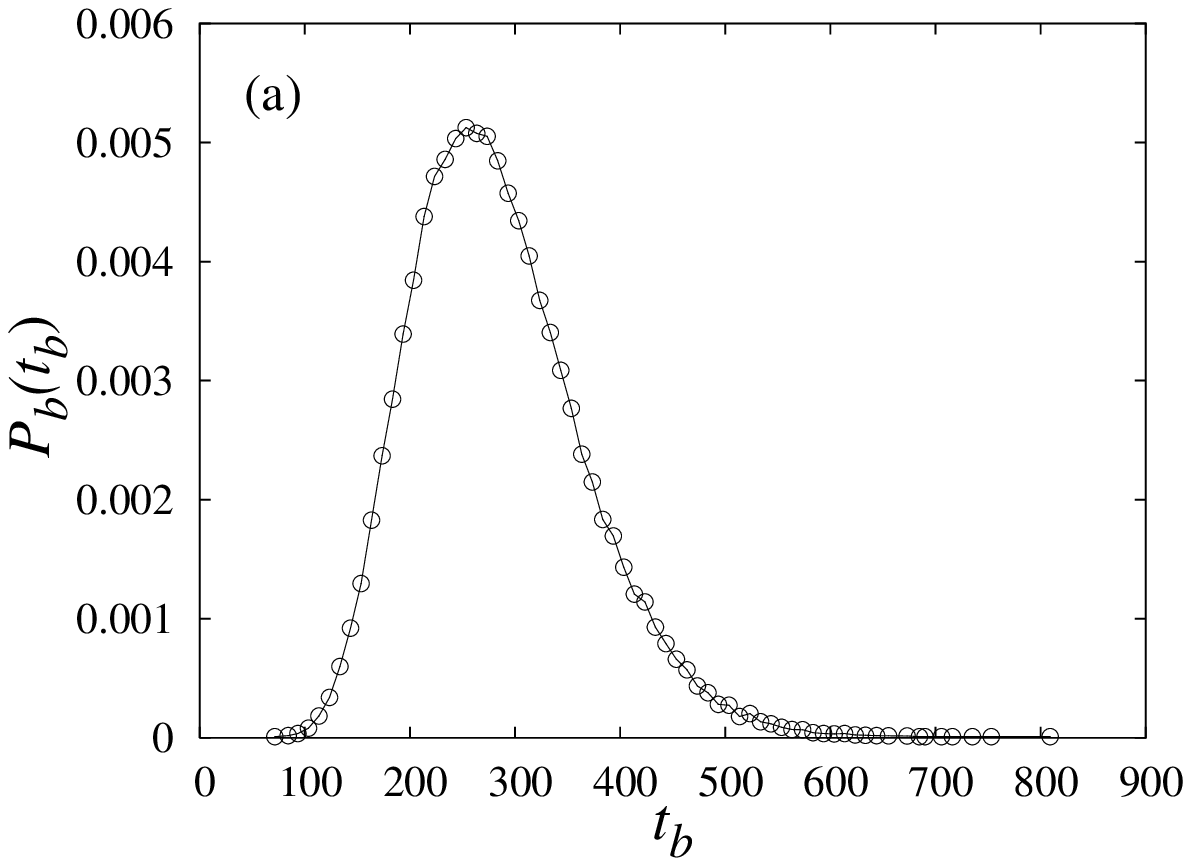}}
\centerline{\epsfxsize=8.5cm \epsfbox{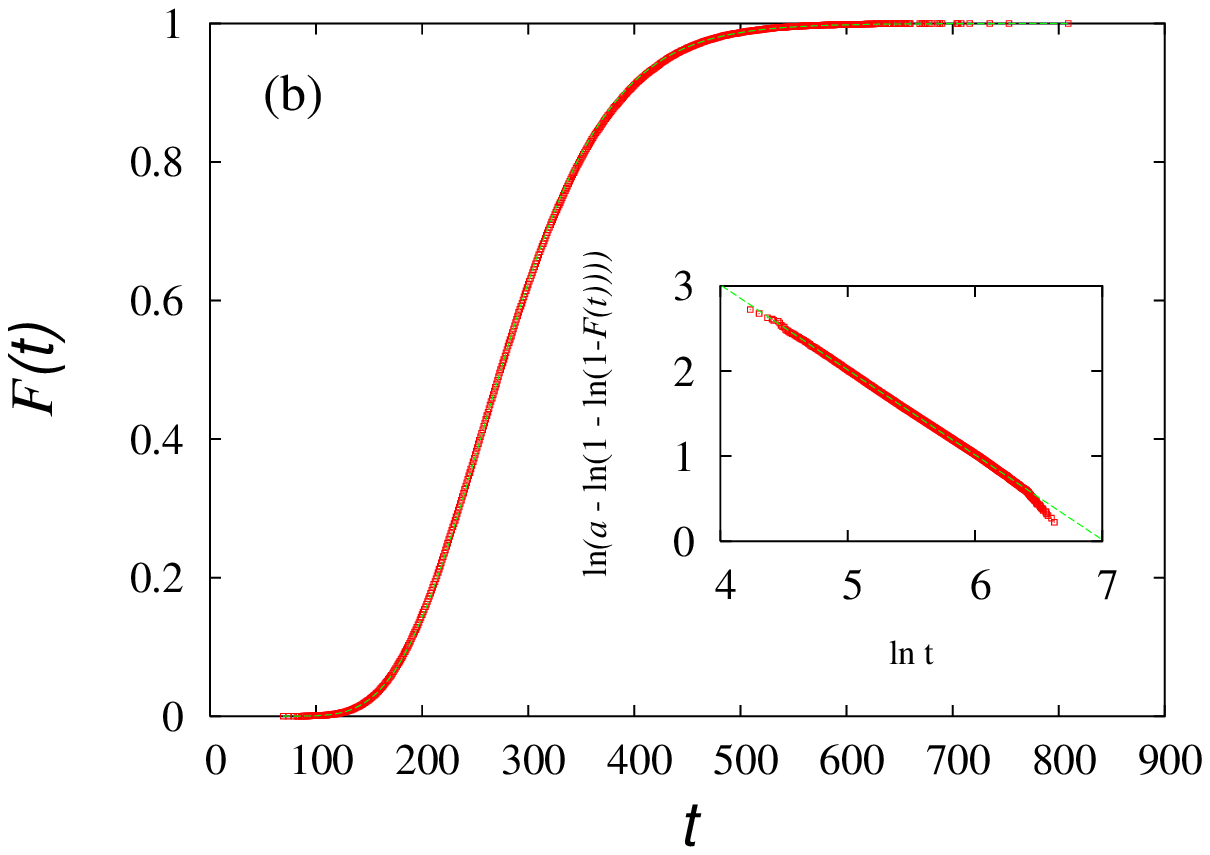}}
\caption{(color online) (a)
Breaking time distribution $P_b(t_b)$ for $\sigma$=13 on 
linear scales. (b) Probability of failure at time $t$ or less,
and $F(t)$ as a function of $t$. Inset: Determining whether $F(t)$ 
fits to the double exponential function.} 
\label{critical_time_distribution}
\end{figure}

\begin{figure}
\centerline{\epsfxsize=8.5cm \epsfbox{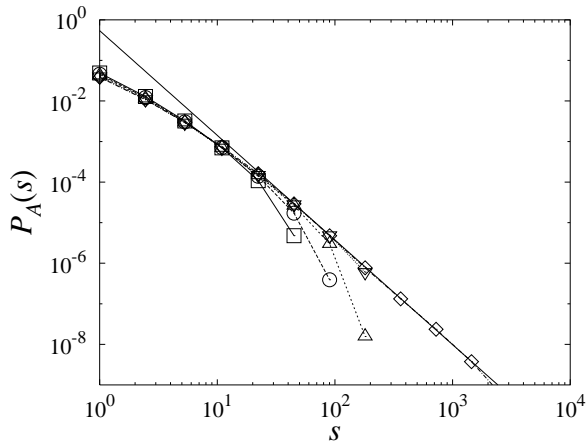}}
\caption{Avalanche size distribution $P_A(s)$ for various $\sigma$
= 12 ({$\Box$}), 13 ({\Large$\circ$}), 14 ({$\triangle$}), 15
({\large$\triangledown$}), and 16 ({\Large$\diamond$}) on double
logarithmic scales. The slope of the linear fit (solid line) is 
$-2.6$. The data are averaged over $10^5$ configurations for
$\sigma=12,13,14$ and 15, but over $55000$ configurations for
$\sigma=16$.} \label{avalanche_size_distribution}
\end{figure}

\begin{figure}
\centerline{\epsfxsize=8.5cm \epsfbox{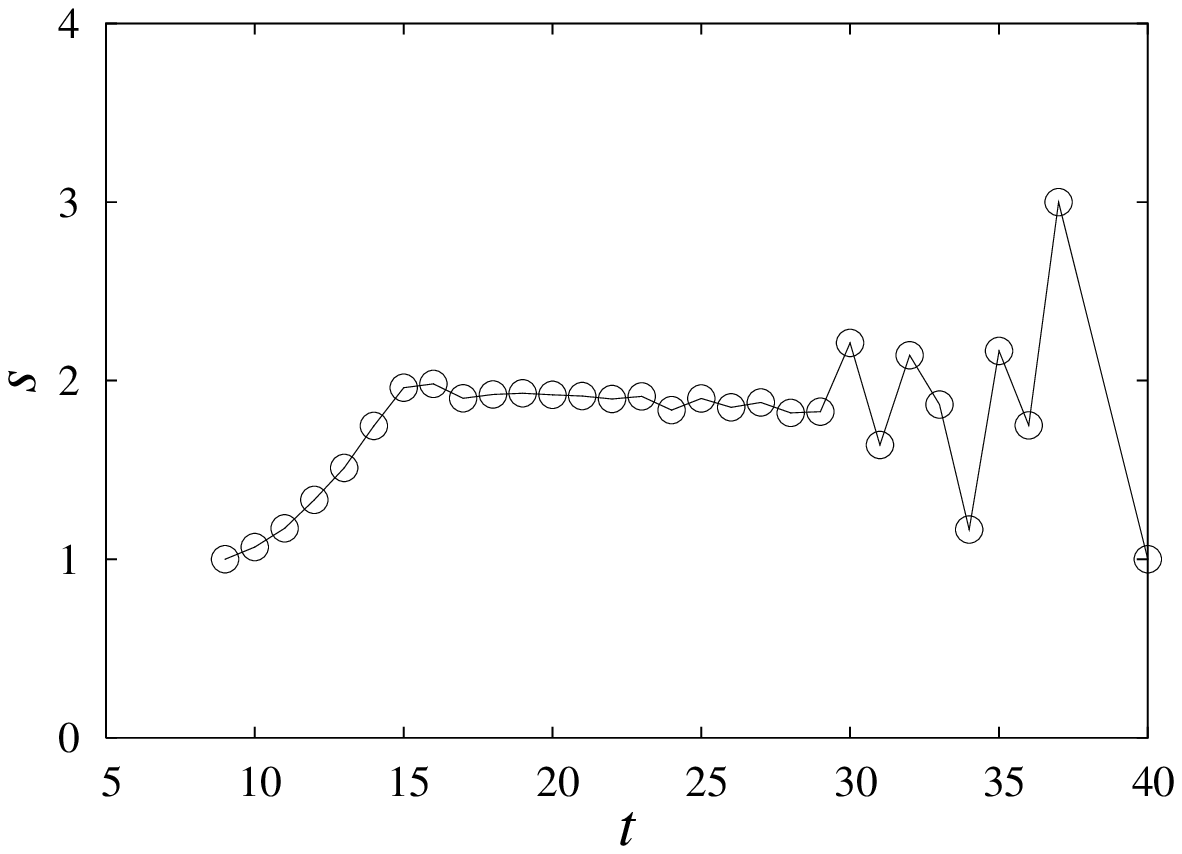}}
\caption{Average avalanche size $\langle s \rangle$ as a function
of time for $\sigma=16$. } \label{average_avalanche_size}
\end{figure}

Using the model, we perform numerical simulations for various values of
$\sigma$. First we considered the breaking time as a function of the
capacity $\sigma$. As shown in Fig.~2, we observe that that the mean 
breaking time $\langle t_b \rangle$ averaged over different ensembles 
increases at a higher rate than the double exponential function. 
This result implies that a small increment in the capacity of each node
significantly enhances the tolerance of the system. That is
because the overload is shared globally. Since the breaking time 
increases very rapidly, it is difficult to perform numerical 
simulations for large values of $\sigma$.

The breaking time fluctuates considerably in
Fig.2, wherein the error bar increases with increasing $\sigma$.
Explicit values are listed in Table 1. Thus, it would be
meaningful to study the distribution of the breaking time since it 
is a relatively sensitive probe of the underlying breaking
mechanism, which is often used as a tool for reliability analysis.
It is conventional to consider the probability $F(t)$ that a
network fails at a time $t$, or less. Then it is expressed 
as $F(t)=\int_0^t P_b(t_b)dt_b$. 
In Fig.3(a), we plot this distribution numerically,
and observe that it fits well to the double exponential form,
\begin{equation}
F(t)=1-\exp[-\exp(a-bt^{-r})],
\end{equation}
which is one of the known functions in extreme value
statistics. The constants $a$, $b$ and $r$ depend on $\sigma$. 
Their values are listed in Table I. The double-exponential 
functional form appears in mechanical failure problems in 
the ductile regime~\cite{ductile}. This functional form
is different from the Weibull distribution that occurs in the brittle
regime and fiber bundle model~\cite{d-hkim}

Next, we study the avalanche size distribution in which the
avalanche size represents the number of nodes that fail successively. 
This distribution follows the power-law,
\begin{equation}
P_A(s)\sim s^{-\tau},
\end{equation}
with the exponent $\tau\approx 2.6(1)$
(Fig.~\ref{avalanche_size_distribution}). The exponent value is
close to that obtained from the fiber bundle model. We determine the
mean avalanche size as a function of triggering time. As shown in
Fig.~5, the mean avalanche size is independent of the triggering time
except in the small $t$ regime.

The exponent of the avalanche size distribution of the toy model is
close to the one obtained in the fiber bundle model in SF
networks~\cite{moreno2002}. In the original one-dimensional 
fiber-bundle model, an external force is evenly distributed to all the
nodes in the network. For $\sigma_i$ of node $i$, a threshold 
value is assigned against failure. When the load is larger 
than the threshold, the node fails irreversibly and its load is 
equally distributed to its remaining nearest neighbors. This
process continues until no failure occurs. Thereafter the process is
repeated by applying a bigger force to the remaining nodes. The
study of the fiber bundle model has been extended to the case of 
complex networks. When a critical force is applied, the avalanche
size distribution follows a power law. Recently Kim {\it et
al.}~\cite{d-hkim} investigated the fiber bundle problem by 
considering various complex networks such as the Erd\H{o}s and 
R\'enyi random network~\cite{er}, the small-world network~\cite{ws} and the
SF network. Their result indicates that the patterns of the avalanche
dynamics occurring in such complex networks are almost the same due 
to the small-world property: The effect of the local load sharing rule is
negligible in complex networks. Therefore the exponent of the
avalanche size distribution reduces to the mean field value, i.e.,
$\tau={5/2}$.

As shown in Fig.5, the avalanche size is independent of time. 
This suggests ineffective correlation between the degrees and 
the entire system is considered to be homogeneous.
Moreover, the avalanche dynamics proceeds globally by the rewiring
dynamics. Thus, the exponent of the avalanche size distribution of
the toy model reduces to the mean field value. Although the toy 
model has not been applied to real world systems, the dynamic rule 
of the rewiring edges in the model is rather unique; it 
may reflect the adaptive behavior of each individual (node)
in the event of a failure. In such a case, the adaptive behavior
does not relieve itself, rather it may lead to other successive failures.
This phenomenon can often be observed in complex systems. For
example, when a city encounters a shortage of electric power, 
the current power grid system is designed to compensate instantly 
by drawing power from neighboring cities. This may cause another 
blackout or a cascading blackout throughout the country. In order to 
prevent such avalanche dynamics induced by the adaptive activity, 
failure should be localized.

This work is supported by the KRF Grant MOEHRD (R14-2002-059-010000-0) 
funded by the Korean government.
%%%%%%%%%%%%%%%%%%%%%%%%%%%%%%%%%%%%%%%%%%%%%%%%%%%%%%%%%%%%%%%%%%%%%%%%%%%%%%
%
% Bibliography
%
%%%%%%%%%%%%%%%%%%%%%%%%%%%%%%%%%%%%%%%%%%%%%%%%%%%%%%%%%%%%%%%%%%%%%%%%%%%%%%

%\end{multicols}
\end{document}